\documentclass[%
 reprint,
 amsmath,amssymb,
 aps,
prx,
]{revtex4-2}

\usepackage{caption}
\usepackage{subcaption}
\usepackage{graphicx}
\usepackage{hyperref}

\usepackage{changes}
\setauthormarkup{}
\setdeletedmarkup{}
\definechangesauthor[name={Nelson}, color=teal]{NL}
\definechangesauthor[name={Carolina}, color=magenta]{CA}
\definechangesauthor[name={Alban}, color=orange]{AS}

\begin{document}

\title{Demonstration of a 10-metre-long discharge plasma source \\  for plasma wakefield acceleration}

\author{C. Amoedo$^{1,2}$}
\email{carolina.amoedo@cern.ch}
\author{N. Lopes$^2$}
\email{nelson.lopes@tecnico.ulisboa.pt}
\author{N. E. Torrado$^{1,2}$}
\author{F. Silva$^3$}
\author{P. Muggli$^{1,4}$}
\author{L. Verra $^{1}$}
\thanks{Present Address: INFN Laboratori Nazionali di Frascati, Frascati, Italy}
\author{M. Turner$^1$}
\author{G. Zevi Della Porta$^1$}
\author{M. Bergamaschi$^1$}
\author{A. Clairembaud$^4$}
\author{J. Mezger$^4$}
\author{F. Pannell$^5$}
\author{N. Z. van Gils$^{1,6}$}
\author{E. Gschwendtner$^1$}
\author{A. Sublet$^1$}
\affiliation{\\}
\collaboration{AWAKE Collaboration}

\author{
C.C. Ahdida$^{1}$,
Y. Alekajbaf$^{7}$,
O. Apsimon$^{8,9}$,
R. Apsimon$^{9,10}$,
T. Bachmann$^{1}$,
C. Badiali$^{2}$,
M. Baquero$^{11}$,
E. Belli$^{1,12}$,
A. Boccardi$^{1}$,
T. Bogey$^{1}$,
S. Burger$^{1}$,
P.N. Burrows$^{12}$,
B. Buttensch\"{o}n$^{13}$,
A. Caldwell$^{4}$,
M. Chung$^{14}$,
C.C. Cobo$^{15}$,
D.A. Cooke$^{5}$,
D. Dancila$^{7}$,
C. Davut$^{8,9}$,
G. Demeter$^{16}$,
A.C. Dexter$^{9,10}$,
S. Doebert$^{2}$,
A. Eager$^{1}$,
D. Easton$^{17}$,
B. Elward$^{18}$,
J. Farmer$^{4}$,
R. Fonseca$^{19,2}$,
I. Furno$^{11}$,
A. Gerbershagen$^{6,12}$,
D. Ghosal$^{9,20}$,
E. Granados$^{1}$,
J. Gregory$^{9,10}$,
O. Grulke$^{13,21}$,
E. Guran$^{1}$,
D. Harryman$^{1}$,
M. Hibberd$^{8,9}$,
H. Jaworska$^{22}$,
P. Karataev$^{23}$,
R. Karimov$^{11}$,
M.A. Kedves$^{16}$,
F. Kraus$^{24}$,
M. Krupa$^{1}$,
T. Lefevre$^{1}$,
K. Lotov,
J. McGunigal$^{8,9}$,
M. Moreira$^{1}$,
B. Moser$^{1}$,
Z. Najmudin$^{15}$,
S. Norman$^{8,9}$,
N. Okhotnikov,
A. Omoumi$^{1}$,
C. Pakuza$^{1}$,
A. Pardons$^{1}$,
J. Pisani$^{17}$,
A. Pukhov$^{22}$,
L. Ranc$^{4}$,
R. Rossel$^{1}$,
H. Saberi$^{8,9}$,
M.V. Santos$^{1}$,
O. Schmitz$^{18}$,
F. Sharmin$^{18}$,
L. Silva$^{2}$,
B. Spear$^{12}$,
L. Stant$^{1}$,
C. Stollberg$^{11}$,
C. Swain$^{9,20}$,
G. Tenasini$^{1}$,
A. Topaloudis$^{1}$,
P. Tuev,
J. Uncles$^{17}$,
F. Velotti$^{1}$,
J. Vieira$^{2}$,
C. Welsch$^{9,20}$,
T. Wilson$^{22}$,
M. Wing$^{5}$,
J. Wolfenden$^{9,20}$,
B. Woolley$^{1}$,
G. Xia$^{8,9}$,
V. Yarygova,
W. Zhang$^{12}$
}

\affiliation{$^1$CERN, Geneva, Switzerland}
\affiliation{$^2$GoLP/IPFN, Instituto Superior Técnico, Universidade de Lisboa, Lisbon, Portugal}
\affiliation{$^3$INESC-ID, Instituto Superior Técnico, Universidade de Lisboa, Lisbon, Portugal}
\affiliation{$^4$Max Planck Institute for Physics, Munich, Germany}
\affiliation{$^5$University College London, London, United Kingdom}
\affiliation{$^6$PARTREC, UMCG, University of Groningen, Groningen, NL}
\affiliation{$^{7}$ Uppsala University, Uppsala, Sweden}
\affiliation{$^{8}$ University of Manchester, Manchester, United Kingdom}
\affiliation{$^{9}$ Cockcroft Institute, Warrington, United Kingdom}
\affiliation{$^{10}$ Lancaster University, Lancaster, United Kingdom}
\affiliation{$^{11}$ École Polytechnique Fédérale de Lausanne (EPFL), Swiss Plasma Center (SPC), Lausanne, Switzerland}
\affiliation{$^{12}$ John Adams Institute for Accelerator Science, University of Oxford, Oxford, United Kingdom }
\affiliation{$^{13}$ Max Planck Institute for Plasma Physics, 17491 Greifswald, Germany}
\affiliation{$^{14}$ POSTECH, Pohang, Republic of Korea}
\affiliation{$^{15}$ Imperial College London, London, United Kingdom}
\affiliation{$^{16}$ HUN-REN Wigner Research Centre for Physics, Budapest, Hungary}
\affiliation{$^{17}$ GWA, Cambridge, United Kingdom}
\affiliation{$^{18}$ University of Wisconsin–Madison, Madison, WI, USA}
\affiliation{$^{19}$ DCTI/ISCTE, Instituto Universitário de Lisboa, Lisbon, Portugal}
\affiliation{$^{20}$ University of Liverpool, Liverpool, United Kingdom}
\affiliation{$^{21}$ Technical University of Denmark, Kongens Lyngby, Denmark}
\affiliation{$^{22}$ Heinrich Heine University Düsseldorf, Düsseldorf, Germany}
\affiliation{$^{23}$ Royal Holloway, University of London, Egham, United Kingdom}
\affiliation{$^{24}$ Universität Bonn, Bonn, Germany}


\begin{abstract}
The AWAKE experiment at CERN uses high-energy proton bunches to drive plasma wakefields capable of accelerating electron bunches up to 
hundreds of GeV in a single plasma. 
Realising this potential requires a plasma source that can extend to tens of metres while maintaining high longitudinal uniformity and reproducibility.
Towards this goal, we
report on the first experimental 
characterisation and beam-based validation
of a 10-metre-long pulsed-DC discharge plasma source (DPS) 
used as an alternative to the laser-ionised rubidium vapour plasma source (VPS).
The DPS reliably delivered
plasma densities in three noble gases (He, Ar, Xe)
at electron densities between 10$^{14}$ and 10$^{15}$ cm$^{-3}$.
Proton bunch self-modulation 
was
observed with 
all three gases, and the resulting modulation frequency 
was
used as a beam-based diagnostic of the plasma density, showing good agreement with laboratory interferometry measurements. The event-to-event modulation frequency was demonstrated to be reproducible to within 0.63\%, comparable to that of the VPS. These results establish
the DPS as a 
valid alternative plasma source for beam-driven wakefield experiments, and support its further development towards longer plasma lengths. The present design is tailored to AWAKE requirements, but the underlying discharge concept provides a route to long, reproducible plasma sources with straightforward density control through discharge–beam timing for other beam–plasma experiments.

\end{abstract}


\maketitle
\section{\label{sec:intro}Introduction}
Plasma-based wakefield acceleration driven by intense laser pulses~\cite{tajima1979laser} or particle bunches~\cite{Chen1985} enables accelerating gradients up to three orders of magnitude higher than those achievable with conventional radio-frequency (RF) structures. However, scaling plasma-based accelerators to energies relevant for high-energy physics remains a central challenge \cite{adolphsen2022european}. Concepts for high-energy machines typically rely on staging of multiple acceleration modules to compensate for the limited stored energy in laser pulses or electron bunches. Each module generally ranges from a few centimetres to approximately one metre in length \cite{hogan2016electron, picksley2020meter, biagioni2025technical, loisch2026kilohertz}, and many successive stages are required to reach TeV-scale beam energies. However, such staging has not yet been demonstrated experimentally and presents significant challenges~\cite{lindstrom2021staging}, especially in the preservation of beam quality and alignment between stages. An alternative is to use a driver carrying sufficient energy to sustain the accelerating interaction over much longer distances~\cite{caldwell2009proton,li2026numerical}. This approach shifts the scalability challenge towards the development of longer plasma sources that can extend the acceleration length within a single stage~\cite{muggli2020physics}. Longer plasma sources are also foreseen for applications including free-electron lasers~\cite{biagioni2025technical}, plasma injectors for storage rings~\cite{farmer2026electron}, and future high-energy colliders~\cite{foster2025halhf,caldwell2025alive}, with specific plasma requirements depending on the driver and application.

The AWAKE experiment at CERN investigates the alternative of using high-energy proton bunches from the Super Proton Synchrotron (SPS) to drive wakefields \cite{gschwendtner2022awake}. With proton energies of 400 GeV and total bunch energies exceeding 19 kJ, a single SPS bunch can drive plasma wakefields over hundreds of metres, making it possible to accelerate electrons to energies of up to hundreds of GeV within a single plasma~\cite{lotov2021plasma}. Using a 10~m laser-ionised rubidium vapour plasma source (VPS)~\cite{oz2014novel}, AWAKE has already demonstrated acceleration of externally injected electrons to 2~GeV~\cite{awake}. To resonantly drive the plasma wakefields, the initially long SPS proton bunch (rms length of $\sim$ 5 cm) must undergo self-modulation (SM), which transforms it into a train of microbunches resonant on the plasma wavelength ($\lambda_{pe}= c/f_{pe}\approx1-3$ mm)~\cite{Kumar2010Self-modulationPlasmas}. The modulation frequency ($f_{mod}$) corresponds to the plasma frequency $f_{pe}$, given by  
\begin{equation}
    f_{pe} = \frac{1}{2\pi} \sqrt{\frac{n_{pe} e^2}{m_e \varepsilon_0}} \quad,
    \label{eq:fmod}
\end{equation}  
where $n_{pe}$ is the plasma electron density, $e$ and $m_e$ the electron charge and mass, respectively, and $ \varepsilon_0 $ the vacuum permittivity. Since $f_{mod}$ is uniquely determined by the plasma electron density, fluctuations in $n_{pe}$ directly affect the resonant excitation of the wakefields and the microbunch periodicity. To preserve this resonant interaction between the microbunch train and the plasma wakefields, variations of the plasma wavelength must remain below $\lambda_{pe}/8$ along the plasma~\cite{lotov2013effect}, which in this case corresponds to a relative plasma density variation below 0.25\% over the plasma length. Maintaining this level of longitudinal density uniformity over tens of metres therefore places particularly stringent demands on long plasma-source technology.

Additionally, the need to maintain a stable and reproducible wakefield phase also places constraints on the development of the self-modulation process. Currently, the laser pulse that ionises the rubidium vapour provides a relativistic ionisation front (RIF) that seeds SM and controls the wakefield phase~\cite{batsch2021transition}. Seeding with an electron bunch has also been demonstrated~\cite{verra2022controlled}, allowing the SM process to be seeded independently of the plasma-generation method (\textit{e.g.} even in a pre-ionised plasma). This independence is particularly relevant to future AWAKE operation, which will use two plasmas: the first dedicated to optimisation of proton bunch self-modulation and the second tailored to electron acceleration~\cite{muggli2020physics,gschwendtner2022awake} over plasma lengths of tens to hundreds of metres. 
While the VPS provides the required longitudinal density uniformity and reproducibility~\cite{batsch2018interferometer} and has been highly successful for studies of proton bunch self-modulation over 10~m~\cite{awake,batsch2021transition,verra2022controlled}, scaling it to much longer lengths is ultimately limited by the ionising laser pulse and its delivery to a few tens of metres in length~\cite{gschwendtner2022awake}.

\begin{figure*}[t!]
    \centering
    \vspace{-5mm}
    \includegraphics[width=.61\textwidth]{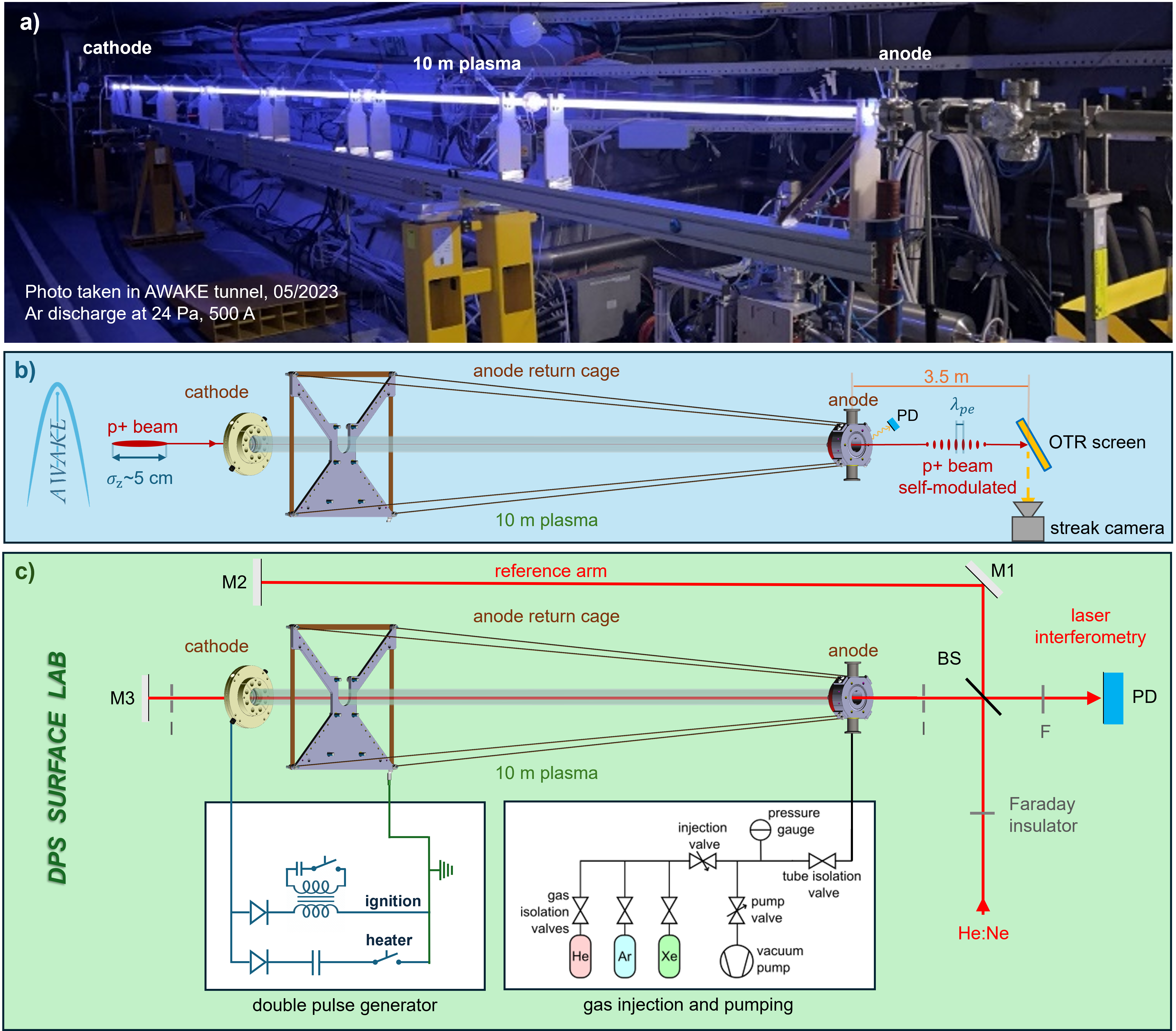}
    \caption{\textbf{(a)} Photograph and \textbf{(b)} schematic of the 10-metre-long DPS in the AWAKE experiment, including the OTR–streak-camera diagnostic \cite{batsch2021transition}. \textbf{(c)} Laboratory configuration showing the Michelson interferometer \cite{amoedo2026lpaw}, the gas-injection system, and a simplified schematic of the double-pulse generator \cite{torrado2023double}, showing the step-up transformer ignition and capacitor heater modules. In \textbf{(c)}: M1–M3 (mirrors), I (irises), BS (beam splitter), F (filter), PD (photodiode).\vspace{-4mm}}
    \label{fig:dps-setup}
\end{figure*}

Alternative plasma-generation concepts capable of extension to greater lengths are therefore required, and AWAKE is currently evaluating two alternative concepts: the Helicon Plasma Source (HPS), based on magnetised RF plasma ~\cite{buttenschon2018high, stollberg2024first, zepp2024direct, granetzny2023preference}, and the Discharge Plasma Source (DPS) described in this work (Fig. \ref{fig:dps-setup}). In addition to length scalability, plasma sources suitable for future AWAKE experiments must provide electron densities in the range of $10^{14}$ to $10^{15}\,\mathrm{cm^{-3}}$ while satisfying the uniformity and reproducibility requirements described above~\cite{muggli2020physics,gschwendtner2022awake}.

The DPS is a double-pulse DC discharge concept originally proposed by Instituto Superior Técnico and Imperial College London~\cite{torrado2023double, lpaw2025}. The double-pulse scheme enables reproducible plasma generation with nanosecond-level timing jitter~\cite{torrado2023double}. The length of the source can be extended by joining multiple adjacent plasma modules, using shared cathodes and anodes~\cite{torrado2024}. Initial studies have demonstrated current equalisation between coupled plasma sections using passive magnetic-balancing modules~\cite{torrado2024}, supporting the length scalability of the DPS. In this paper, we present the first integration of a 10-metre-long DPS into the AWAKE experiment. 
The DPS was first characterised in a dedicated laboratory at CERN using time-resolved interferometry, providing measurements of the temporal evolution of the longitudinally integrated plasma density in three noble gases: xenon (Xe), argon (Ar), and helium (He). It was then installed in the AWAKE experiment, where the self-modulation instability of the proton bunch was observed in plasmas of these gases. Because this work is particularly relevant for the second plasma (\textit{accelerator}), no SM seeding mechanism was used, and the study remained with the SMI process. The direct dependence of $f_{mod}$ on $n_{pe}$ (Eq.~\ref{eq:fmod}) enabled SM to serve as a beam-based diagnostic of the line-integrated plasma electron density~\cite{adli2019experimental}. The successful integration of the DPS into AWAKE also enabled dedicated studies of ion-mass effects on SM \cite{turner2024experimental} and of beam-driven filamentation \cite{verra2024filamentation}, illustrating its versatility as an alternative to the VPS. 

Both interferometry and the beam-based diagnostic provide line-integrated information on the plasma density and therefore cannot resolve its longitudinal uniformity. They do, however, enable measurements of event-to-event reproducibility, which were compared with that achieved by the VPS, an important consideration for the future use of the DPS.

This paper is organised as follows: section \ref{sec:dps} introduces the experimental setup together with the laboratory plasma density characterisation based on time-resolved interferometry; section \ref{sec:smi} presents the beam-based plasma density measurements obtained from the modulation frequency of the proton bunch, and section \ref{sec:discussion} discusses the implications of these results for future scalable plasma sources in AWAKE. 

\section{\label{sec:dps}Experimental setup}
\begin{figure*}[t!]
    \centering
    \vspace{-3mm}
    \begin{subfigure}{.39\linewidth}
            \caption{}
        \centering
        \includegraphics[width=\linewidth]{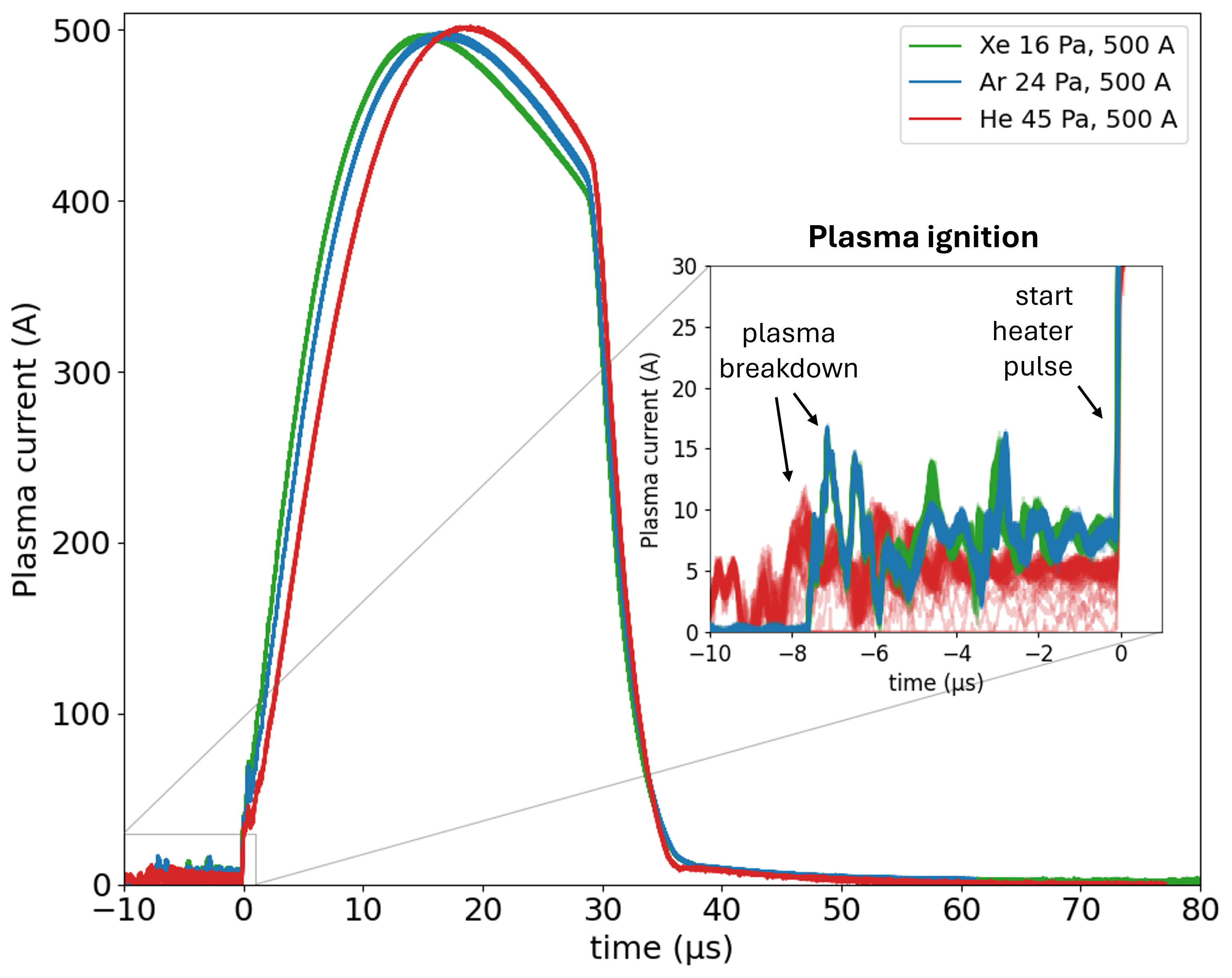}
        \label{fig:current_gases}
    \end{subfigure}
    \hspace{8mm}
    \begin{subfigure}{.39\linewidth}
    \caption{}
        \centering
        \includegraphics[width=\linewidth]{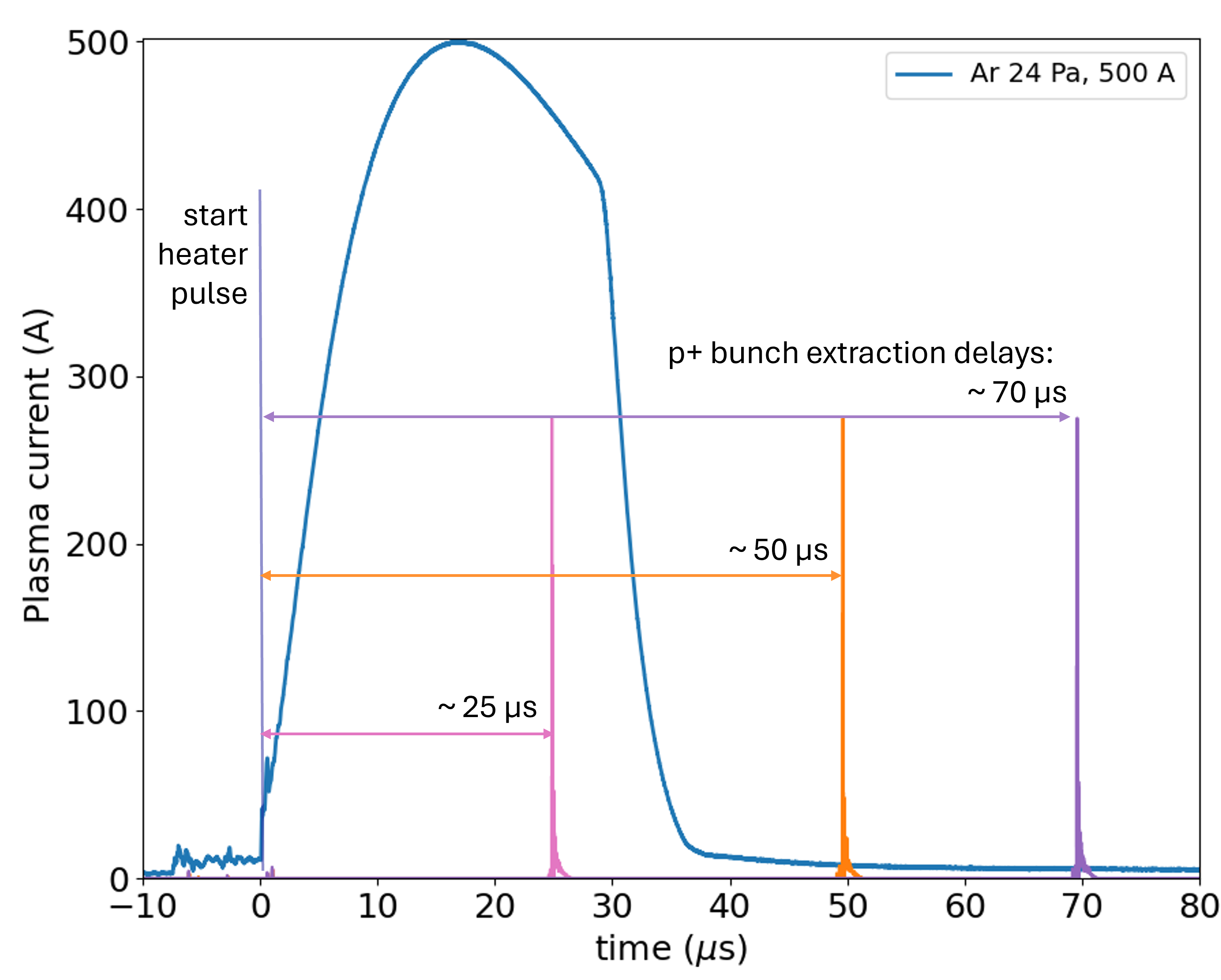}
        \label{fig:arrival-protons}
    \end{subfigure}
    \vspace{-4mm}
    \caption{\textbf{(a):} Discharge current evolution for  Xe (16 Pa), Ar (24 Pa), and He (45 Pa) at a peak current of approximately 500 A. For each gas, 200 consecutive discharge traces are overlaid. \textit{Inset:} Close-up of the plasma ignition, showing the event-to-event jitter associated with breakdown and the start of the heater pulse.  \textbf{(b):} Timing of the proton bunch extractions relative to the start of the heater pulse ($t=0$) for the reference Ar discharge (24 Pa, 500 A). The three extraction delays $\simeq(25$, 50, and 70 $\mu$s), measured by the photodiode signal, are indicated in pink, orange, and purple, shown with arbitrarily scaled amplitudes for visual clarity.\vspace{-4mm}}
\end{figure*}
The experiments were performed using a 10-metre-long DPS, consisting of a glass tube with a 26-mm inner diameter and electrodes located at each end (Fig. \ref{fig:dps-setup}). This prototype was first characterised in a surface laboratory (Fig. \ref{fig:dps-setup} c) and subsequently installed in the AWAKE experiment (Fig. \ref{fig:dps-setup} a,b), where it was operated during a three-week proton-beam run \cite{gschwendtner2024jacow}.

The source was operated with Xe, Ar and He at pressures of 16, 24, and 45 Pa, respectively. Gas pressure was controlled through regulated gas injection and pumping downstream of the anode (Fig. \ref{fig:dps-setup} c) and varied by less than $\pm$1\% from event to event. These operating conditions were selected to provide access to the plasma density range of $10^{14}$ to $10^{15}\,\text{cm}^{-3}$ required for the AWAKE studies reported here and in \cite{turner2024experimental,verra2024filamentation}.

The plasma was generated using the double-pulse generator shown in Fig.~\ref{fig:dps-setup}c, consisting of a high-voltage ignition stage based on a step-up flyback transformer and a capacitor discharge heater stage \cite{torrado2023double}. The ignition pulse (15--25~kV) initiated the discharge, while the subsequent $\approx30~\mu$s heater pulse ($\sim 6$kV) drove currents up to 500~A and set the plasma electron density. The short heater duration was chosen 
to avoid non-uniformities such as striations, which typically develop in DC discharges at times $>200~\mu$s~\cite{anders2024glows}. 

Figure \ref{fig:current_gases} shows the discharge current waveforms for the three gases. 
The primary difference between the discharges in the three gases is the reproducibility of the plasma breakdown following the ignition pulse. Helium exhibits a larger variation in the breakdown time than Xe and Ar, as shown in the inset of Fig.~\ref{fig:current_gases}. 
The approximately 10~$\mu$s delay between the ignition and heater pulses is sufficiently long to accommodate the different breakdown times of the three gases, allowing the heater pulse to be triggered at a fixed time. The start of the heater pulse is defined as $t=0$ and provides a reproducible timing reference for setting the delay between plasma generation and proton-bunch extraction, as illustrated in Fig.~\ref{fig:arrival-protons}. Under the operating conditions used here, the three gases reach comparable peak discharge currents of approximately 500~A, corresponding to an average current density of $\sim100~\mathrm{A/cm^2}$, with an event-to-event reproducibility better than 1\% of the peak current~\cite{torrado2024}. 

\begin{figure*}[t!]
    \centering
    \vspace{-4mm}
    \includegraphics[width=.75\linewidth]{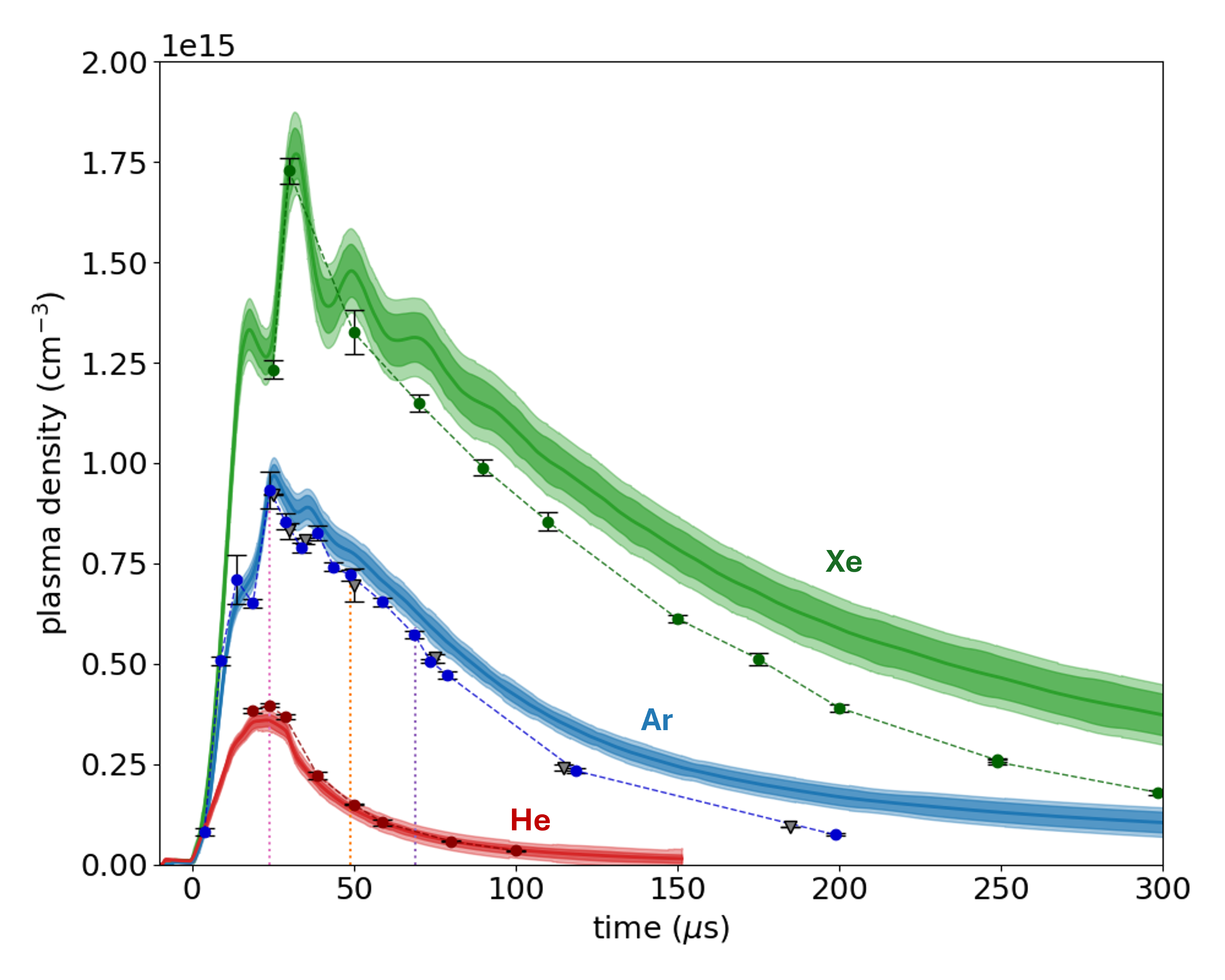}
\vspace{-4mm}
    \caption{Plasma density evolution for Xe (red), Ar (blue), and He (green), obtained from interferometry measurements (solid lines and shaded regions) and from self-modulation frequency measurements (points with error bars), discussed in Section~\ref{sec:smi}. Grey triangles indicate the independent set of Ar measurements acquired after more than 20,000 additional discharge cycles. Vertical lines show the delays $t\simeq(25, 50, 70 )\mu$s introduced in Fig. \ref{fig:arrival-protons} and \ref{fig:dft}.\vspace{-4mm}}
    \label{fig:smi_gases}
\end{figure*}

Despite their similar current evolution, the plasma density differs significantly across gases, reflecting the different first-ionisation energies (Xe: 12.1~eV, Ar: 15.8~eV, He: 24.6~eV) and electron-impact ionisation cross-sections \cite{rapp1965total}. Figure~\ref{fig:smi_gases} shows the average plasma density evolution for 500 A discharge current in Xe, Ar and He discharges reconstructed from 200 consecutive events acquired at 0.1~Hz, matching the SPS repetition rate used in AWAKE. The density evolution was measured using a Michelson interferometer, in which a probe beam (approx. 2 mm diameter) traversed the plasma twice, corresponding to a total averaging length of 20~m (Fig. \ref{fig:dps-setup}c). The plasma-induced phase shift is proportional to the line-integrated electron density, and the density reconstruction procedure is described in detail in \cite{amoedo2026lpaw}. In Fig.~\ref{fig:smi_gases}, the solid lines represent the average density evolution over the analysed discharges. For all three gases, the event-to-event variation (darker shaded region) is smaller than the uncertainty associated with the density reconstruction (lighter shaded region), indicating that the intrinsic discharge reproducibility exceeds the precision of the interferometric diagnostic~\cite{amoedo2026lpaw}.

With the same 500 A current setpoint, the highest density is reached in Xe, followed by Ar and He, with average peak plasma densities of $(1.78,0.87,0.36)\times10^{15}\,\mathrm{cm^{-3}}$, respectively. Ar discharges reach the full density range required for AWAKE operation ($10^{14}$ to $10^{15}$ cm$^{-3}$)~\cite{muggli2018awake}, while Xe offers a straightforward route to achieve higher plasma densities under similar discharge conditions.\par Helium reaches lower plasma densities and presents the greatest challenges during plasma ignition~\cite{torrado2024} and shows larger event-to-event ignition jitter (inset Fig.~\ref{fig:current_gases}); however, its low ion mass makes it particularly interesting for studies of ion-motion effects on the SM process, as explored in \cite{turner2024experimental}.  

Figure~\ref{fig:smi_gases} shows that all three discharges generate a rapid increase in plasma density, followed by a gradual decay due to recombination. In He, this evolution is smooth, whereas the Xe and Ar discharges exhibit density oscillations around the peak density, visible as successive local maxima and minima. The larger amplitude of these oscillations observed in Xe is consistent with the trend previously observed in Ar discharges, where they became more pronounced with increasing plasma density \cite{amoedo2026lpaw}. 
These density oscillations may be associated with
the development of radial plasma dynamics following the rapid ionisation and heating of the discharge core.
Although a detailed investigation of these oscillations is beyond the scope of this work, supplementary radial measurements \cite{amoedo2026phd} are consistent with the development of a time-dependent radial plasma structure. The possible implications of this radial evolution for DPS operation are discussed in Section \ref{sec:discussion}.

Finally, the reproducible evolution of the plasma density (Fig.~\ref{fig:smi_gases}) provides a direct mapping between the plasma density and the time elapsed since the start of the heater pulse (defined as $t=0$), allowing different plasma conditions to be reproducibly accessed by adjusting the proton bunch extraction timing.

\section{\label{sec:smi}Self-modulation instability measurements}
Following the laboratory characterisation, the DPS was used in AWAKE (Fig. \ref{fig:dps-setup}a) with SPS proton bunches, containing approximately $3\times10^{11}$ protons and with an rms bunch length of $\sim$190 ps and transverse size of 200 $\mu$m. Figure~\ref{fig:arrival-protons} illustrates three representative extraction delays used during the AWAKE measurements, corresponding to different plasma densities along the evolution shown in Fig.~\ref{fig:smi_gases}. The timing between the discharge and the SPS proton bunch was monitored using a fast photodiode located downstream of the plasma source, which detects light generated by the interaction of the proton bunch with beamline elements.

Time-resolved images of the proton bunch current density distribution after propagation in the plasma were measured using the standard diagnostic system based on optical transition radiation (OTR) and a streak camera \cite{batsch2021transition, verra2022controlled,adli2019experimental, turner2024experimental, nechaeva2024hosing}. The OTR generated when the proton bunch crosses a metallic foil 3.5 m downstream of the plasma (Fig. \ref{fig:dps-setup}b)
reproduces
the spatio-temporal bunch current density. The streak camera records a 73 ps time window sampled by 512 pixels, and with an overall temporal resolution of approximately 1 ps, sufficient to resolve the microbunching structure. The modulation frequency extracted from these images is used to infer the plasma electron density (Eq.~\ref{eq:fmod}).

\begin{figure*}[t!]
\vspace{-2.5mm}
    \centering
    \begin{subfigure}[t]{.42\linewidth}
    \caption{\label{fig:waterfall}}
        \centering
        \includegraphics[width=\linewidth]{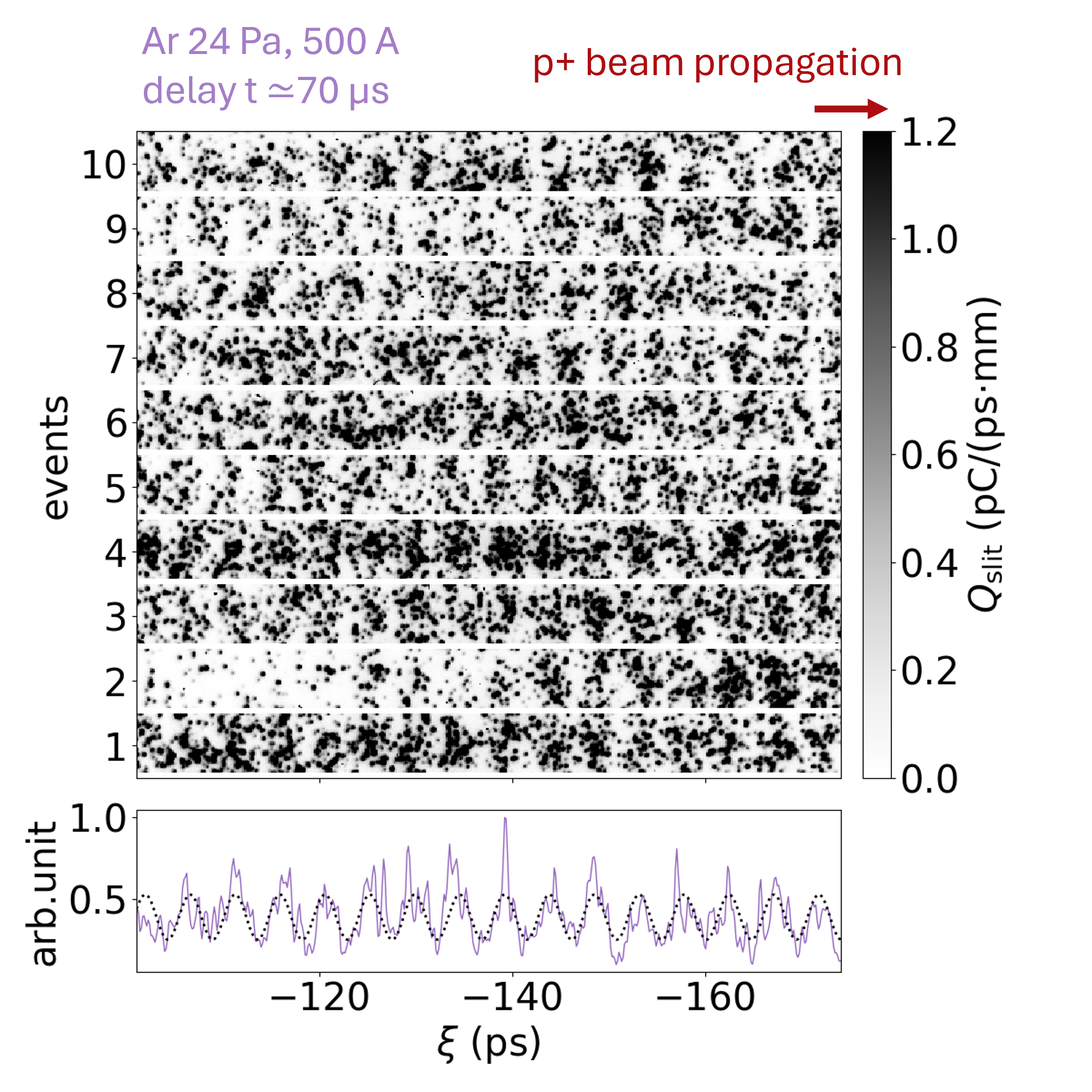}
    \end{subfigure}
    \begin{subfigure}[t]{.42\linewidth}
    \vspace{5mm}
    \caption{\label{fig:dft}}
        \centering
        \includegraphics[width=\linewidth]{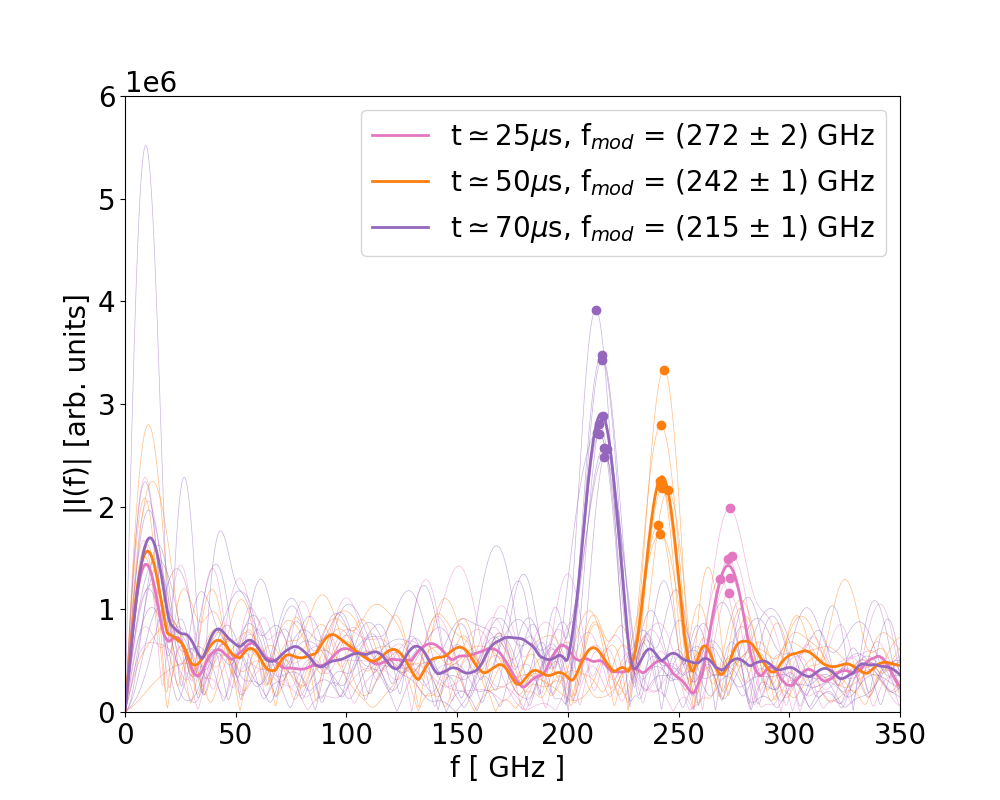}
    \end{subfigure}
    \vspace{-4mm}
    \caption{\textbf{(a)} Time-resolved images of 10 individual SPS proton bunch events of (Ar 24 Pa, 500 A, $t\simeq70\mu$s), recorded on a 73-picosecond timescale. The colour scale is saturated to enhance the visibility of the microbunch structure. The longitudinal profile of event 1 (bottom) and a sinusoidal fit using the average modulation frequency (215 GHz) are shown below. \textbf{(b)} DFT power spectra of individual events (thin lines) and their corresponding average spectra (thick lines) for the three delays $t\simeq(25, 50, 70 \mu$s) introduced in Fig. \ref{fig:arrival-protons}. Dots indicate the peak $f_{mod}$ extracted for each individual spectrum.\vspace{-4mm} \label{fig:smi-10-events}}
\end{figure*}

Figure~\ref{fig:waterfall} shows 10 representative streak camera images obtained for an Ar discharge (24 Pa, 500 A) with the proton bunch arriving approximately 70~$\mu$s after the start of the heater pulse, corresponding to the latest extraction delay shown in Fig.~\ref{fig:arrival-protons}. The periodic microbunch structure is clearly resolved in these images, providing direct evidence of the development of SMI in the DPS. Similar modulation was observed for all extraction delays and in all three gases investigated in this work. 
Since the instability was unseeded, its development varied from event to event, leading to fluctuations in the phase (the time of the first microbunch in the image) and in the OTR light intensity. For instance, in event 2 in Fig.~\ref{fig:waterfall}, the modulation appears 
to have developed earlier in the propagation distance, or more strongly than in the other cases, with a significant 
reduction of charge observed later along the bunch, consistent with stronger transverse fields. 

To determine the modulation frequency, the streak camera images were first integrated in the transverse (spatial) direction, yielding a one-dimensional longitudinal bunch profile for each event (as shown at the bottom of Fig.~\ref{fig:waterfall} for event 1). The modulation frequency was then extracted using discrete Fourier transform (DFT) analysis of these profiles, following the method described in~\cite{batsch2021transition}. Prior to the DFT, the longitudinal profiles were zero-padded by a factor of 50, reducing the frequency-grid spacing to approximately 0.35~GHz, which allows a more precise interpolation of the spectral peak. The modulation frequency is identified as this dominant peak, whose amplitude must exceed twice the noise level, estimated from the mean spectral amplitude between 50 and 500~GHz. Despite the event-to-event differences seen in Fig.~\ref{fig:waterfall}, the corresponding DFT power spectra for all events at the same delay exhibit a consistent $f_{mod}$, as shown in Fig.~\ref{fig:dft}. Figure~\ref{fig:dft} shows the spectra corresponding to the extraction delays introduced in Fig.~\ref{fig:arrival-protons} and the average modulation frequencies found are $f_{mod} = (272 \pm 2,\; 242 \pm 1,\; 215 \pm 1)\,\text{GHz}$ for $t \simeq (25, 50, 70)\,\mu\text{s}$, respectively. Using Eq.~\ref{eq:fmod}, these correspond to plasma densities of $n_{pe} = (9.33 \pm 0.14,\; 7.21 \pm 0.06,\; 5.72 \pm 0.05)\times 10^{14}\,\text{cm}^{-3}$. Here, the quoted frequencies correspond to the average of ten consecutive events, while the uncertainties represent the largest difference found between individual measurements and the mean.

The same procedure was repeated for the Xe, Ar, and He discharges at different extraction delays along the temporal evolution of the plasma, as shown in Fig.~\ref{fig:smi_gases}. The resulting plasma densities inferred from $f_{mod}$ (Eq.~\ref{eq:fmod}) are presented as points with error bars. For all three gases, the inferred densities agree well with the independent interferometry measurements. In particular, with Ar, where the temporal scan was performed with smaller delay steps, the beam-based measurements also reproduce the oscillatory features observed with interferometry. Comparing the three gases, the agreement is strongest for He, while systematic deviations between the two measurements are observed for Ar and Xe discharges during the recombination phase (\textit{i.e.}, late in time). These differences for Ar and Xe are discussed further in Sec.~\ref{sec:discussion}. 

\begin{figure}[t!]
    \centering
    \vspace{-2mm}
    \includegraphics[width=.95\linewidth]{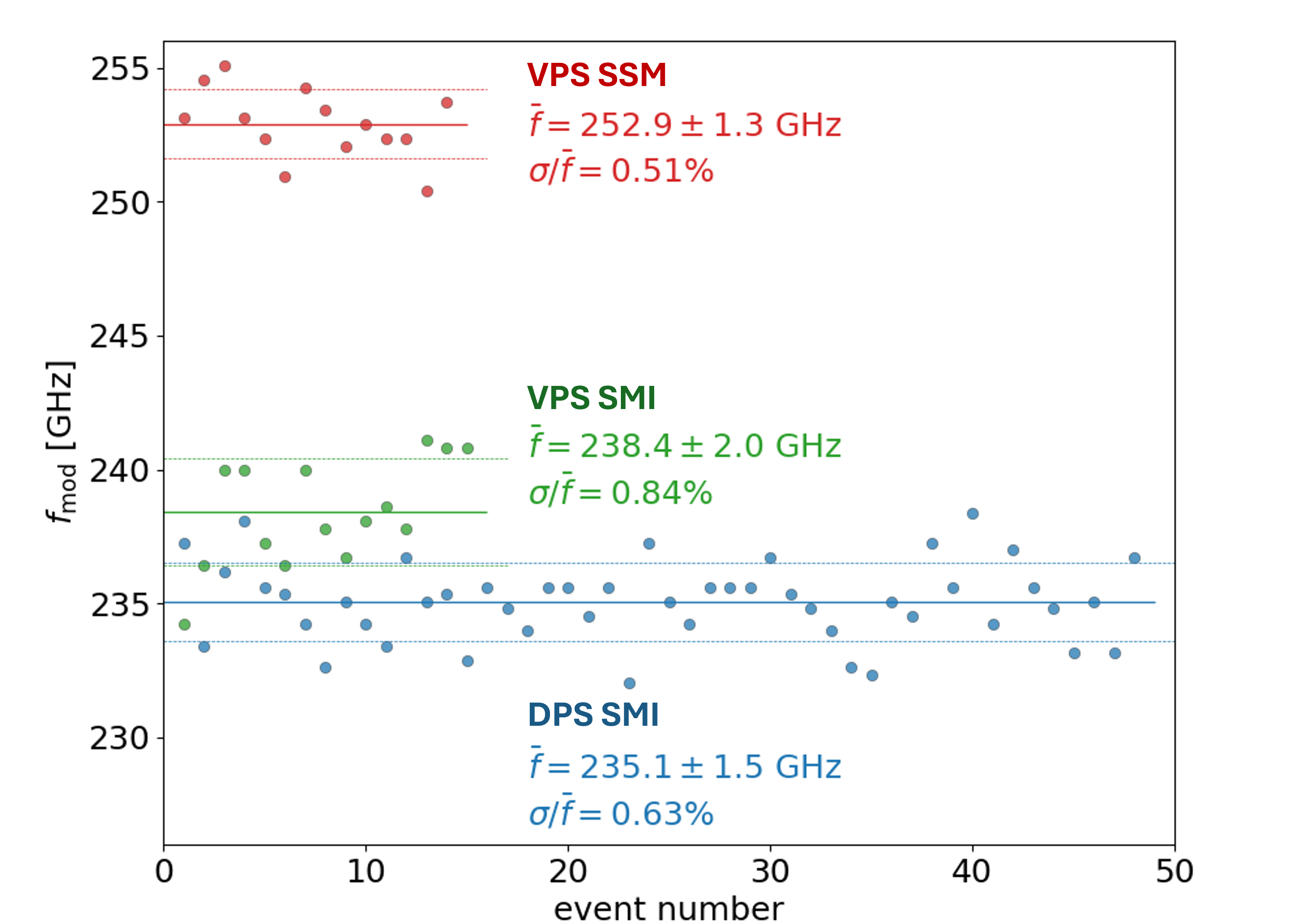}
    \caption{Event-to-event variation of the measured $f_{mod}$ in the DPS (blue dots, Ar discharges, at 24 Pa, 500 A, delay $\simeq$ 50 $\mu$s, $n_{pe}=6.86\times10^{14}\,\text{cm}^{-3}$) and the VPS under unseeded (SMI, green dots, $n_{pe}=7.1\times10^{14}\,\text{cm}^{-3}$) and seeded SM (SSM, red dots, $n_{pe}=7.95\times10^{14}\,\text{cm}^{-3}$). The mean frequency (continuous line) and relative standard deviation ($\sigma/f$, dashed line) are indicated for each dataset. The modulation frequency reproducibility measured with the DPS lies within the range observed for the VPS under both seeded and unseeded SM conditions.\vspace{-4mm}}
    \label{fig:shot_DPS_Rb}
\end{figure}

To assess the reproducibility of the modulation frequency under fixed plasma conditions, 48 consecutive events were recorded at around $n_e \simeq 7\times10^{14}\text{cm}^{-3}$ in Ar discharges with 24~Pa and 500~A peak current, corresponding to the proton bunch 
extraction approximately equal to $50~\mu$s after the start of the heater pulse. The plasma density presented here corresponds to the nominal AWAKE plasma density~\cite{muggli2018awake}, but similar results were obtained at the other plasma densities investigated. The modulation frequencies extracted from these events are shown in blue in Fig.~\ref{fig:shot_DPS_Rb}. These are distributed within $(235.1 \pm 1.5)$ GHz, corresponding to $n_{pe}=(6.86 \pm 0.09)\times10^{14}\,\text{cm}^{-3}$. 

To directly compare the performance of the DPS with the vapour plasma source currently used in AWAKE, Fig.~\ref{fig:shot_DPS_Rb} also shows two subsets of 14 SM events recorded in the VPS that were analysed using the same frequency-extraction procedure. The first dataset corresponds to SMI (green in Fig.~\ref{fig:shot_DPS_Rb}), obtained by positioning the RIF 1 ns ahead of the proton bunch centre, such that the laser pulse serves only to create the plasma and does not seed SMI. The second dataset corresponds to seeded self-modulation (SSM, in red in Fig.~\ref{fig:shot_DPS_Rb}), with the RIF located at the centre of the bunch \cite{batsch2018interferometer}. The two datasets were recorded with uniform vapour densities of $7.1\times10^{14}\,\text{cm}^{-3}$ and $7.95\times10^{14}\,\text{cm}^{-3}$, respectively \cite{oz2014novel,batsch2018interferometer}, under proton bunch conditions comparable to those used in the present study.

As shown in Fig.~\ref{fig:shot_DPS_Rb}, the DPS results exhibit a relative standard deviation of $f_{mod}$ of 0.63\%, comparable to the values measured with the VPS for both SMI (0.72\%) and SSM (0.51\%). These values are also comparable to those previously reported in \cite{batsch2018interferometer} (Supplementary Material) for lower plasma densities. The similar $f_{mod}$ spread measured for SMI in the two plasma sources demonstrates that the DPS performs at the same level as the VPS used in AWAKE. Furthermore, the comparable $f_{mod}$ variations measured for SMI and SSM in the VPS indicate that the observed spread is essentially independent of the seeding conditions, suggesting that any difference between the two regimes is below the precision of the present diagnostic and frequency-extraction procedure.

These event-to-event $f_{mod}$ variations correspond to relative plasma density variations of approximately 1--1.4\%, since $f_{\mathrm{mod}}\propto\sqrt{n_{pe}}$. However, this value is substantially larger than the independently measured density reproducibility of the VPS ($\leq0.2\%$) \cite{oz2014novel}, which suggests that the measured $f_{mod}$ spread is dominated by the diagnostic and frequency-extraction procedure rather than by intrinsic plasma density fluctuations. During the 48-event DPS dataset, the discharge operating conditions were themselves highly reproducible, with relative variations of 0.29\% in the gas pressure and 0.88\% in the peak discharge current. Although these quantities are operational parameters rather than direct measurements of the plasma density, their stability provides further evidence of the reproducibility of the discharge. Moreover, the observed fluctuations do not represent the fundamental reproducibility limit of the source. The gas pressure stability could be further improved by actively stabilising the gas temperature, while the discharge current reproducibility is expected to benefit from improved matching and precision of the pulsed-power components. 

In addition, an independent set of SM measurements, shown by the grey triangles in Fig.~\ref{fig:smi_gases}, was acquired with Ar after more than 20,000 additional discharge cycles (three weeks of beam-time~\cite{torrado2023double}) and showed the same relation between extraction delay and modulation frequency, within the experimental uncertainties, as the measurements acquired at the beginning of the run. These measurements demonstrate both event-to-event reproducibility and reliability over extended operation, which further supports the conclusion that, within the precision of the present diagnostic, the performance of the DPS is comparable to that of the VPS, strengthening its case as an alternative plasma source for future AWAKE operation.

\section{Discussion\label{sec:discussion}}

The different levels of agreement between the plasma densities inferred from the self-modulation frequency and those measured independently by interferometry (Fig.~\ref{fig:smi_gases}) for the three discharge gases provide complementary insight into the temporal evolution of the plasma.
These differences are most pronounced in Ar and Xe discharges during the recombination phase, when the plasma densities inferred from the SM measurements are systematically lower than those measured by interferometry by approximately 9\%. As discussed in Section~\ref{sec:dps}, these are also the plasmas that exhibit the strongest temporal density oscillations. 
If these oscillations arise from a radial expansion of the plasma, as suggested in \cite{amoedo2026lpaw}, diagnostics probing different transverse regions may not observe the same density evolution. The interferometer averages the plasma density over the $\sim$2~mm probe-beam diameter, whereas the proton bunch samples a much smaller region with a transverse size of approximately 200~$\mu$m. Consequently, the comparison becomes increasingly sensitive to radial density gradients and to small misalignments between the plasma and the diagnostics. In addition, the cumulative nature of the interferometric phase-retrieval procedure \cite{amoedo2026lpaw} amplifies any bias introduced by the onset of radial density gradients on the densities inferred during recombination.
These effects together provide a plausible explanation for the systematic and larger differences observed during the recombination phase of the discharge and motivate the combined use of local and transverse plasma diagnostics, such as Thomson scattering \cite{stollberg2024first,amoedo2026TS}, plasma imaging and transverse interferometry~\cite{granetzny2025innovative}, to further resolve the spatio-temporal evolution of the discharge.


The ability to operate the DPS with different noble gases also enabled the experimental investigation of ion-mass effects on proton bunch self-modulation \cite{turner2024experimental}. Lighter gases exhibit stronger ion motion, which reduces the wakefield amplitude towards the back of the proton bunch \cite{turner2024experimental,vieira2014self,walter2025ion}. This effect does not influence the modulation frequency measurements presented here, since the frequency is extracted from the front of the bunch, before significant ion motion develops. For plasma-wakefield acceleration, however, ion motion places a practical constraint on the choice of gas, favouring ion masses greater than or equal to that of Ar \cite{vieira2014self,walter2025ion}, which is the baseline gas for the DPS. Additionally, the successful operation of the DPS with Xe provides a practical route to higher plasma densities while satisfying this requirement.

The delay-density mapping shown here facilitates the integration of the DPS into the two-plasma configuration envisaged for future AWAKE operation \cite{gschwendtner2022awake}, allowing the plasma density to be optimised to that of the upstream plasma by adjusting only the discharge timing from event to event. Together with the length scalability of the discharge concept, this opens a path towards the long acceleration plasmas required for future AWAKE experiments.

The results presented also define the next steps in the development of long discharge plasmas. While this work demonstrates reproducible plasma generation and beam-based validation over 10~m, plasmas for controlled acceleration will require quantitative measurements of longitudinal plasma density uniformity at the $\sim0.25\%$ level needed for stable wakefield phase. The comparison between interferometry and beam-based measurements highlights the need for diagnostics capable of resolving local plasma properties \cite{stollberg2024first}, including radial and longitudinal density profiles, rather than relying exclusively on longitudinally integrated measurements. In parallel, ongoing developments on coupling adjacent discharges \cite{torrado2024} provide a route towards extending the discharge concept beyond the present 10-metre-long prototype, enabling longer plasmas for acceleration for AWAKE \cite{gschwendtner2022awake} and other plasma-wakefield accelerators \cite{caldwell2025alive,farmer2026electron, foster2025halhf}. The modular configuration may also provide a route to controlled longitudinal density steps between adjacent discharge sections, extending its functionality beyond simple length scaling.

\section{\label{sec:conclusions}Conclusions}

This work presents the experimental characterisation and beam-based validation of a 10-metre-long pulsed-DC discharge plasma source designed to provide length-scalable plasmas for the AWAKE experiment in its programme towards demonstrating particle acceleration from tens to potentially hundreds of GeV. The plasma source was integrated into the experiment and reliably delivered plasmas in three noble gases (He, Ar, and Xe) with plasma electron densities between 10$^{14}$ and 10$^{15}$~cm$^{-3}$.

Proton bunch self-modulation was observed with all three gases, and the modulation frequency was used as a beam-based diagnostic of the plasma density, showing good agreement with laboratory interferometry measurements. An event-to-event modulation frequency variation of 0.63\% was demonstrated, comparable to that of the VPS under seeded and unseeded SM conditions, which represents a notable result given that the DPS operates in the unseeded self-modulation regime. Measurements repeated after more than 20000 discharge cycles recovered the same delay–density relation within the experimental uncertainties, demonstrating the reliability of the source over extended operation times.
The flexibility of operation of the DPS, including discharges in different noble gases and simple density tunability through discharge–beam timing control, enabled dedicated studies of ion mass effects \cite{turner2024experimental} and beam-driven filamentation \cite{verra2024filamentation}, expanding the experimental reach of AWAKE.

Future work will focus on extending the DPS to longer plasma lengths by joining discharge modules using shared cathodes and anodes, and on demonstrating the sub-0.25\% density uniformity and reproducibility required for electron acceleration, which will require both further engineering development of the source and improved plasma diagnostics.

These results establish the DPS as a valid alternative plasma source for beam-driven wakefield experiments. While the present design is tailored to AWAKE requirements, the same principles are transferable to other plasma wakefield experiments, with ongoing studies already considering adaptations of discharge plasma sources to different density and repetition-rate requirements~\cite{caldwell2025alive,farmer2026electron}. More broadly, the development of long plasma sources with high density uniformity, reproducibility, tunability, and stability is common to several proposed plasma-based accelerator and collider concepts \cite{adolphsen2022european,foster2025halhf,gessner2025design,cros2025contribution}, providing direct synergies with the DPS development presented here.

\begin{acknowledgments}
This work was supported by Fundação para a Ciência e Tecnologia – Portugal (Grant No. 2024.00263.CERN \url{https://doi.org/10.54499/2024.00263.CERN}).
The work in AWAKE was supported in part by STFC (AWAKE-UK, Cockcroft Institute core, John Adams Institute core, and UCL consolidated grants) and the National Research Foundation of Korea (Grant No. NRF-2016R1A5A1013277 and NRF-2020R1A2C1010835). Support of the Wigner Datacenter Cloud facility through the Awakelaser project is acknowledged. M. W. acknowledges the support of DESY, Hamburg and UW Madison acknowledges support by NSF Award No. PHY-1903316. The AWAKE collaboration acknowledges the SPS team for their excellent proton delivery.
\end{acknowledgments}

\bibliographystyle{elsarticle-num}
\bibliography{2026_DPS_SMI}

@article{tajima1979laser,
  title={Laser electron accelerator},
  author={Tajima, Toshiki and Dawson, John M},
  journal={Physical Review Letters},
  volume={43},
  number={4},
  pages={267},
  year={1979},
  publisher={APS},
  doi={10.1103/PhysRevLett.43.267}
}

@article{hogan2016electron,
  title={Energy doubling of 42 {GeV} electrons in a metre-scale plasma wakefield accelerator},
  author={Blumenfeld, Ian and Clayton, Christopher E and Decker, Franz-Josef and Hogan, Mark J and Huang, Chengkun and Ischebeck, Rasmus and Iverson, Richard and Joshi, Chandrashekhar and Katsouleas, Thomas and Kirby, Neil and others},
  journal={Nature},
  volume={445},
  number={7129},
  pages={741--744},
  year={2007},
  publisher={Nature Publishing Group UK London},
  doi={10.1038/nature05538}
}

@article{lindstrom2021staging,
  title={Staging of plasma-wakefield accelerators},
  author={Lindstr{\o}m, Carl A},
  journal={Physical Review Accelerators and Beams},
  volume={24},
  number={1},
  pages={014801},
  year={2021},
  publisher={APS},
  doi={10.1103/PhysRevAccelBeams.24.014801}
}

@article{lotov2021plasma,
  title={Plasma wakefield acceleration beyond the dephasing limit with 400 {GeV} proton driver},
  author={Lotov, KV and Tuev, PV},
  journal={Plasma Physics and Controlled Fusion},
  volume={63},
  number={12},
  pages={125027},
  year={2021},
  publisher={IOP Publishing},
  doi={10.1088/1361-6587/ac349a}
}

@article{vieira2014self,
  title={Self-modulation instability of ultra-relativistic particle bunches with finite rise times},
  author={Vieira, J and Amorim, LD and Fang, Y and Mori, WB and Muggli, P and Silva, LO},
  journal={Plasma Physics and Controlled Fusion},
  volume={56},
  number={8},
  pages={084014},
  year={2014},
  publisher={IOP Publishing},
  doi={10.1088/0741-3335/56/8/084014}
}

@article{adli2019experimental,
  title={Experimental observation of proton bunch modulation in a plasma at varying plasma densities},
  author={{AWAKE Collaboration}},
  journal={Physical Review Letters},
  volume={122},
  number={5},
  pages={054802},
  year={2019},
  publisher={APS},
  doi={10.1103/PhysRevLett.122.054802}
}

@article{batsch2021transition,
  title={Transition between instability and seeded self-modulation of a relativistic particle bunch in plasma},
  author={Batsch, Fabian and {et al. (AWAKE Collaboration)}},
  journal={Physical Review Letters},
  volume={126},
  number={16},
  pages={164802},
  year={2021},
  publisher={APS},
  doi={10.1103/PhysRevLett.126.164802}
}

@article{oz2014novel,
  title={A novel {Rb} vapor plasma source for plasma wakefield accelerators},
  author={{\"O}z, E and Muggli, P},
  journal={Nuclear Instruments and Methods in Physics Research Section A: Accelerators, Spectrometers, Detectors and Associated Equipment},
  volume={740},
  pages={197--202},
  year={2014},
  publisher={Elsevier},
  doi={10.1016/j.nima.2013.10.093}
}

@article{Chen1985,
author = {Chen, P. and Dawson, J.M. and Huff, Robert W. and Katsouleas, T.},
doi = {10.1103/PhysRevLett.54.693},
issn = {03759601},
journal = {Physical Review Letters},
number = {7},
pages = {693},
title = {{Acceleration of electrons by the interaction of a bunched electron beam with a plasma}},
volume = {54},
year = {1985}
}

@article{awake,
  title={Acceleration of electrons in the plasma wakefield of a proton bunch},
  author={{AWAKE Collaboration}},
  journal={Nature},
  volume={561},
  number={7723},
  pages={363--367},
  year={2018},
  publisher={Nature Publishing Group UK London},
  doi={10.1038/s41586-018-0485-4}
}

@article{caldwell2009proton,
  title={Proton-driven plasma-wakefield acceleration},
  author={Caldwell, Allen and Lotov, Konstantin and Pukhov, Alexander and Simon, Frank},
  journal={Nature Physics},
  volume={5},
  number={5},
  pages={363--367},
  year={2009},
  publisher={Nature Publishing Group UK London},
  doi={10.1038/nphys1248}
}

@article{gschwendtner2022awake,
  title={The {AWAKE} Run 2 programme and beyond},
  author={Gschwendtner, Edda and Lotov, Konstantin and Muggli, Patric and Wing, Matthew and {et al. (AWAKE Collaboration)}},
  journal={Symmetry},
  volume={14},
  number={8},
  pages={1680},
  year={2022},
  publisher={MDPI},
  doi={10.3390/sym14081680}
}

@article{picksley2020meter,
    author = {Miao, B. and Shrock, J. E. and Rockafellow, E. and Sloss, A. J. and Milchberg, H. M.},
    title = {Meter-scale supersonic gas jets for multi-{GeV} laser-plasma accelerators},
    journal = {Review of Scientific Instruments},
    volume = {96},
    number = {4},
    pages = {043003},
    year = {2025},
    month = {04},
    issn = {0034-6748},
    doi = {10.1063/5.0248959}
}

@article{verra2022controlled,
  title={Controlled growth of the self-modulation of a relativistic proton bunch in plasma},
  author={Verra, L and Zevi Della Porta, G and Pucek, J and Nechaeva, T and Wyler, S and Bergamaschi, M and Senes, E and Guran, E and Moody, JT and Kedves, M A and Gschwendtner, E and Muggli, P and {et al. (AWAKE Collaboration)}},
  journal={Physical Review Letters},
  volume={129},
  number={2},
  pages={024802},
  year={2022},
  publisher={APS}, 
  doi={10.1103/PhysRevLett.129.024802}
}

@article{batsch2018interferometer,
  title={Interferometer-based high-accuracy white light measurement of neutral rubidium density and gradient at {AWAKE}},
  author={Batsch, Fabian and Martyanov, Mikhail and Oez, Erdem and Moody, Joshua and Gschwendtner, Edda and Caldwell, Allen and Muggli, Patric},
  journal={Nuclear Instruments and Methods in Physics Research Section A: Accelerators, Spectrometers, Detectors and Associated Equipment},
  volume={909},
  pages={359--363},
  year={2018},
  publisher={Elsevier},
  doi={10.1016/j.nima.2018.02.067}
}

@article{lotov2013effect,
  title={Effect of plasma inhomogeneity on plasma wakefield acceleration driven by long bunches},
  author={Lotov, KV and Pukhov, A and Caldwell, A},
  journal={Physics of Plasmas},
  volume={20},
  number={013102},
  year={2013},
  publisher={AIP Publishing},
  doi={10.1063/1.4773905}
}

@article{stollberg2024first,
  title={First {Thomson} scattering results from {AWAKE}’s helicon plasma source},
  author={Stollberg, Christine and Guittienne, Ph and Karimov, Renat and Sublet, A and Furno, Ivo and Vincent, Benjamin and Andrebe, Yanis and Buttensch{\"o}n, B},
  journal={Plasma Physics and Controlled Fusion},
  volume={66},
  number={11},
  pages={115011},
  year={2024},
  publisher={IOP Publishing},
  doi={10.1088/1361-6587/ad7d36}
}

@article{zepp2024direct,
  title={Direct measurement of the {2D} axisymmetric ionization source rate in a helicon plasma for wakefield particle accelerator applications},
  author={Zepp, M and Granetzny, M and Schmitz, O},
  journal={Physics of Plasmas},
  volume={31},
  number={070704},
  year={2024},
  publisher={AIP Publishing},
  doi={10.1063/5.0211109}
}

@article{granetzny2023preference,
  title={Preference of right-handed whistler modes and helicon discharge directionality due to plasma density gradients},
  author={Granetzny, Marcel and Schmitz, Oliver and Zepp, Michael},
  journal={Physics of Plasmas},
  volume={30},
  number={120701},
  year={2023},
  publisher={AIP Publishing},
  doi={10.1063/5.0173918}
}

@article{buttenschon2018high,
  title={A high power, high density helicon discharge for the plasma wakefield accelerator experiment {AWAKE}},
  author={Buttensch{\"o}n, Birger and Fahrenkamp, Nils and Grulke, Olaf},
  journal={Plasma Physics and Controlled Fusion},
  volume={60},
  number={7},
  pages={075005},
  year={2018},
  publisher={IOP Publishing},
  doi={10.1088/1361-6587/aac13a}
}

@article{verra2024filamentation,
  title={Filamentation of a relativistic proton bunch in plasma},
  author={Verra, L and Amoedo, C and Torrado, N and Clairembaud, A and Mezger, J and Pannell, F and Pucek, J and van Gils, N and Bergamaschi, M and Zevi Della Porta, G and Lopes, N and Sublet, A and  Turner, M and Gschwendtner, E and Muggli, P and {et al. (AWAKE Collaboration)}},
  journal={Physical Review E},
  volume={109},
  number={5},
  pages={055203},
  year={2024},
  publisher={APS},
  doi={10.1103/PhysRevE.109.055203}
}

@article{turner2024experimental,
  title={Experimental observation of motion of ions in a resonantly driven plasma wakefield accelerator},
  author={Turner, M and Walter, E and Amoedo, C and Torrado, N and Lopes, N and Sublet, A and Bergamaschi, M and Pucek, J and Mezger, J and van Gils, N and Verra, L and Zevi Della Porta, G and Farmer, J and Clairembaud, A and Pannell, F and Gschwendtner, E and Muggli, P and  {et al. (AWAKE Collaboration)}},
  journal={Phys. Rev. Lett},
  year={2025},
  publisher={APS},
  volume={134},
  pages={155001},
 doi = {10.1103/PhysRevLett.134.155001}
}

@article{torrado2023double,
  title={Double pulse generator for unipolar discharges in long plasma tubes for the {AWAKE} experiment},
  author={Torrado, Nuno E and Lopes, Nelson C and Silva, J Fernando A and Amoedo, Carolina and Sublet, Alban},
  journal={IEEE Transactions on Plasma Science},
  volume={51},
  number={12},
  pages={3619--3627},
  year={2023},
  publisher={IEEE},
  doi={10.1109/TPS.2023.3337314}
}

@article{torrado2024,
  title={Double pulse generator for AWAKE scalable discharge plasma source},
  author={Torrado, NE and Amoedo, C and Sublet, A and Taborelli, M and Pinto, SF and Silva, JF and Lopes, NC},
  journal = {Journal of Physics: Conference Series},
  volume={3124},
  number={1},
  pages={012021},
  year={2025},
  organization={IOP Publishing},
doi={10.1088/1742-6596/3124/1/012021}
}

@article{lpaw2025,
  title={A metre-scale plasma discharge for plasma wakefield acceleration},
  author={Lopes, Nelson MC and Cobo, Claudia C and Bin Ben Chen, Jian and Jurj, Paul Bogdan and Kennedy, Lewis C and Vanstone, Alexander and Najmudin, Zulfikar},
  journal={Plasma Physics and Controlled Fusion},
  volume={68},
  number={4},
  pages={045046},
  year={2026},
  publisher={IOP Publishing},
doi={10.1088/1361-6587/ae5adc}
}

@article{amoedo2026lpaw,
    title={Time-resolved characterisation of a 10 m discharge plasma for {AWAKE}},
  author={Amoedo, Carolina and Sublet, Alban and Torrado, Nuno and Silva, Fernando A and Lopes, Nelson},
  journal={Plasma Physics and Controlled Fusion},
  volume={68},
  number={6},
  pages={065014},
  year={2026},
  publisher={IOP Publishing}, 
  doi={10.1088/1361-6587/ae72ca}
}

@article{amoedo2026TS,
  title={Local plasma density measurements of a Discharge Plasma Source for {AWAKE} using {Thomson} Scattering},
  author={Amoedo, C and Sublet, A and Santos, M and Stollberg, C and Karimov, R and Lopes, N},
  year={2026},
  journal={in preparation}
}

@article{gschwendtner2024jacow,
  title={Results and plans for run 2 of the advanced proton driven plasma wakefield acceleration experiment AWAKE},
  author={Gschwendtner, Edda},
  journal={JACoW IPAC},
  volume={2024},
  pages={MOPR41},
  year={2024},
  doi={10.18429/JACoW-IPAC2024-MOPR41}
}

@article{anders2024glows,
  title={Glows, arcs, ohmic discharges: An electrode-centered review on discharge modes and the transitions between them},
  author={Anders, Andr{\'e}},
  journal={Applied Physics Reviews},
  volume={11},
  number={031310},
  year={2024},
  publisher={AIP Publishing},
  doi={10.1063/5.0205274}
}

@article{nechaeva2024hosing,
  title={Hosing of a long relativistic particle bunch in plasma},
  author={Nechaeva, Tatiana and Verra, L and Pucek, J and Ranc, L and Bergamaschi, M and Zevi Della Porta, G and Muggli, Patric and {et al. (AWAKE Collaboration)}},
  journal={Physical Review Letters},
  volume={132},
  number={7},
  pages={075001},
  year={2024},
  publisher={APS},
  doi={10.1103/PhysRevLett.132.075001}
}

@article{granetzny2025innovative,
  title={An innovative heterodyne microwave interferometer for plasma density measurements on the {Madison AWAKE} prototype},
  author={Granetzny, Marcel and Elward, Barret and Schmitz, Oliver},
  journal={Review of Scientific Instruments},
  volume={96},
  number={9},
  year={2025},
  publisher={AIP Publishing},
doi={10.1063/5.0271408}
}

@article{adolphsen2022european,
  title={European Strategy for Particle Physics--Accelerator R\&D Roadmap},
  author={Adolphsen, C and Angal-Kalinin, D and Arndt, T and Arnold, Maurice and Assmann, R and Auchmann, B and Aulenbacher, K and Ballarino, A and Baudouy, B and Baudrenghien, P and others},
  journal={arXiv preprint arXiv:2201.07895},
  year={2022},
doi={10.23731/CYRM-2022-001}
}

@article{caldwell2025alive,
  title={Proton-Driven Plasma Wakefield Acceleration for Future HEP Colliders},
  author={Caldwell, Allen and Farmer, John and Lopes, Nelson and Pukhov, Alexander and Willeke, Ferdinand and Wilson, Thomas},
  journal={arXiv preprint },
  year={2025},
doi={10.48550/arXiv.2503.21669}
}

@article{Kumar2010Self-modulationPlasmas,
    title = {{Self-modulation instability of a long proton bunch in plasmas}},
    year = {2010},
    journal = {Physical Review Letters},
    author = {Kumar, Naveen and Pukhov, Alexander and Lotov, Konstantin},
    number = {25},
    pages = {255003},
    volume = {104},
    publisher = {APS},
    doi={10.1103/PhysRevLett.104.255003}
}

@phdthesis{amoedo2026phd,
  author       = {Maria Carolina Amoedo Goncalves},
  title        = {A discharge plasma source for proton-driven plasma wakefield acceleration at CERN},
  school       = {Universidade de Lisboa, Instituto Superior Técnico},
  year         = {2026},
  type         = {PhD thesis},
  doi={10.17181/cpp4v-4r277}
}

@article{farmer2026electron,
  title={An electron injector for the Electron-Ion Collider based on proton-driven plasma wakefield acceleration},
  author={Farmer, JP and Jaworska, H and Caldwell, A and Lopes, N and Pukhov, A and Reichwein, L and Willeke, F and Wing, M},
  journal={arXiv preprint},
  year={2026},
  doi={10.48550/arXiv.2605.07929}
}

@inproceedings{muggli2020physics,
  title={Physics to plan {AWAKE} Run 2},
  author={Muggli, Patric and {for the AWAKE collaboration}},
  booktitle={Journal of Physics: Conference Series},
  volume={1596},
  number={1},
  pages={012008},
  year={2020},
  organization={IOP Publishing},
  doi={10.1088/1742-6596/1596/1/012008}
}

@article{rapp1965total,
  title={Total cross sections for ionization and attachment in gases by electron impact. I. Positive ionization},
  author={Rapp, Donald and Englander-Golden, Paula},
  journal={The Journal of Chemical Physics},
  volume={43},
  number={5},
  pages={1464--1479},
  year={1965},
  publisher={American Institute of Physics},
doi={10.1063/1.1696957}
}

@misc{walter2025ion,
  doi = {10.48550/ARXIV.2512.14476},
  author = {Walter,  Erwin and Farmer,  John P. and Turner,  Marlene and Jenko,  Frank},
  title = {Influence of ion motion in a resonantly driven wakefield accelerator},
  journal = {arXiv preprint},
  year = {2025},
  copyright = {arXiv.org perpetual,  non-exclusive license}
}

@article{muggli2018awake,
  title={{AWAKE} readiness for the study of the seeded self-modulation of a 400 {GeV} proton bunch},
  author={Muggli, Patrick and Adli, Erik and Apsimon, Robert and Asmus, F and Baartman, R and Bachmann, Anna-Maria and Barros Marin, M and Batsch, Fabian and Bauche, Jeremie and Berglyd Olsen, VK and others},
  journal={Plasma Physics and Controlled Fusion},
  volume={60},
  number={1},
  pages={014046},
  year={2018},
  publisher={IOP Publishing},
  doi={10.1088/1361-6587/aa941c}
}

@article{loisch2026kilohertz,
  title={Kilohertz repetition rate capillary discharge pulse modulator with energy recuperation and energy deposition monitoring},
  author={Loisch, Gregor and Kahl, Joachim and Obier, Frank and Teichgr{\"a}ber, Jan Lukas},
  journal={Plasma Physics and Controlled Fusion},
  volume={68},
  number={2},
  pages={025007},
  year={2026},
  publisher={IOP Publishing},
  doi={10.1088/1361-6587/ae3ad6}
}

@article{biagioni2025technical,
  title={Technical status report on plasma components and systems in the context of {EuPRAXIA}},
  author={Biagioni, A and Bourgeois, N and Brandi, F and Cassou, K and Corner, L and Crincoli, L and Cros, B and Dobosz Dufr{\'e}noy, S and Douillet, D and Drobniak, P and others},
  journal={Physics of Plasmas},
  volume={32},
  number={11},
  year={2025},
  publisher={AIP Publishing},
  doi={10.1063/5.0286730}
}

@article{foster2025halhf,
      title={{HALHF}: a hybrid, asymmetric, linear Higgs factory using plasma- and RF-based acceleration}, 
      author={Erik Adli and others},
      year={2025},
      eprint={2503.19880},
      archivePrefix={arXiv},
      primaryClass={physics.acc-ph},
}

@article{gessner2025design,
  title={Design initiative for a 10 Tev pCM wakefield collider},
  author={Gessner, Spencer and Osterhoff, Jens and Lindstr{\o}m, Carl A and Cassou, Kevin and Griso, Simone Pagan and List, Jenny and Adli, Erik and Foster, Brian and Palastro, John and Donegani, Elena and others},
  journal={arXiv preprint arXiv:2503.20214},
  year={2025},
doi={10.48550/arXiv.2503.20214}
}

@article{cros2025contribution,
  title={Contribution of {ALEGRO} to the Update of the European Strategy on Particle Physics},
  author={Cros, Brigitte and Muggli, Patric and Corner, Laura and Farmer, John and Ferarrio, Massimo and Gessner, Spencer and Gizzi, Leo and Gschwendtner, Edda and Hogan, Mark and Hooker, Simon and others},
  journal={arXiv preprint arXiv:2504.01434},
  year={2025},
  doi={10.48550/arXiv.2504.01434}
}

@article{li2026numerical,
  title={Numerical investigations of heavy ion-driven plasma wakefield acceleration},
  author={Li, Jiangdong and Xia, Guoxing and Yang, Jiancheng and Liu, Jie and Zhu, Ruihu and Li, Guangxian},
  journal={Physics of Plasmas},
  volume={33},
  number={3},
  year={2026},
  publisher={AIP Publishing}, 
  doi={10.1063/5.0316747}
}
\end{document}